\documentclass[smallextended]{svjour3}
\usepackage{graphicx}

\smartqed  
\journalname{Transport in Porous Media}
\begin{document}

\title{Relations between Seepage Velocities in Immiscible, Incompressible 
Two-Phase Flow in Porous Media}
\author{Alex Hansen \and Santanu Sinha \and Dick Bedeaux \and Signe
Kjelstrup \and Magnus Aa.\ Gjennestad \and Morten Vassvik}
\date{Received: date / Accepted: date}
\titlerunning{Seepage velocity relations}
\maketitle

\begin{abstract}
Based on thermodynamic considerations we derive a set of equations
relating the seepage velocities of the fluid components in immiscible 
and incompressible two-phase flow in porous media.  They necessitate the
introduction of a new velocity function, the co-moving velocity.  This
velocity function is a characteristic of the porous medium.  Together
with a constitutive relation between the velocities and the driving forces, 
such as the pressure gradient, these equations form a closed set.  We solve 
four versions of the capillary tube model analytically using this theory. 
We test the theory numerically on a network model.     
\end{abstract}

\keywords{Immiscible two-phase flow \and thermodynamics \and seepage velocity \and capillary 
tube model \and network models}

\institute{Alex Hansen \and Magnus Aa.\ Gjennestad \and Morten Vassvik \at
              \address{PoreLab and Department of Physics\\
              Norwegian University of Science and Technology\\
              NO-7491 Trondheim\\
              Norway\\
              \email{Alex.Hansen@ntnu.no}\\}}
\institute{Santanu Sinha \at
              \adress{CSRC\\
              10 West Dongbeiwang Road\\
              Haidian District, Beijing 100193\\
              China\\
              \email{santanu@csrc.ac.cn}\\}}
\institute{Dick Bedeaux \and Signe Kjelstrup \at
              \address{PoreLab and Department of Chemistry\\
              Norwegian University of Science and Technology\\
              NO-7491 Trondheim\\
              Norway\\
              \email{Signe.Kjelstrup@ntnu.no}\\}} 


\section{Introduction}

\label{intro}

The simultaneous flow of immiscible fluids through porous media has been
studied for a long time \cite{b72}. It is a problem that lies at the heart
of many important geophysical and industrial processes. Often, the length
scales in the problem span numerous decades; from the pores measured in
micrometers to reservoir scales measured in kilometers. At the largest
scales, the porous medium is treated as a continuum governed by effective
equations that encode the physics at the pore scale.

The problem of tying the pore scale physics together with the effective
description at large scale is the upscaling problem. In 1936, Wycoff and
Botset proposed a generalization of the Darcy equation to immiscible
two-phase flow \cite{wb36}. It is instructive to reread Wycoff and Botset's
article. This is where the concept of relative permeability is introduced.
The paper is eighty years old and yet it is still remarkably modern.
Capillary pressure was first considered by Richards as early as 1931 \cite{r31}. 
In 1940, Leverett combined capillary pressure with the concept of
relative permeability, and the framework which dominates all later practical  
analyses of immiscible multiphase flow in porous media was in place \cite{l40}.

However, other theories exist \cite%
{lsd81,gh89,hg90,hg93a,hg93b,h98,hb00,gm05,h06a,h06b,h06c,hd10,nbh11,dhh12,bt13,habgo15,%
h15,gsd16}. These theories are typically based on a number of detailed 
assumptions concerning the porous medium and concerning the physics involved. 

It is the aim of this paper to present a new theory for flow in porous media
that is based on thermodynamic considerations.   In the same way as 
Buckley and Leverett's analysis based on the conservation of the mass of the 
fluids in the porous medium led to their Buckley-Leverett equation \cite{bl42}, 
the thermodynamical considerations presented here leads to a set of equations
that are general in that they transcend detailed assumptions about the
flow.  This is in contrast to the relative permeability equations
that rely on a number of specific physical assumptions.     

The theory we present focuses on one aspect of thermodynamic theory. We see
it as a starting point for a more general analysis based on non-equilibrium 
thermodynamics \cite{kp98,kb08,kbjg17} which combines conservation laws with 
the laws of thermodynamics. The structure of the theory, already as it is presented
here, is reminiscent of the structure of thermodynamics itself: we have a number 
of variables that are related through general thermodynamic principles leaving 
an equation of state to account for the detailed physics of the problem. To our
knowledge, the analysis we present here has no predecessor.  However, the 
framework in which it resides is that originally laid out by Gray and 
Hassanizadeh \cite{gh89}, Hassanizadeh and Gray \cite{hg90,hg93a,hg93b}, and Gray 
and Miller \cite{gm05}.

The theory we present concerns relations between the flow rates, or
equivalently, the seepage velocities of the immiscible fluids.  We do not
discuss relations between the seepage velocities and the driving forces that
create them, such as the pressure gradient.  In this sense, we are presenting 
a ``kinetic" theory of immiscible two-phase flow in porous media, leaving 
out the ``dynamic" aspects which will be addressed elsewhere.  
Our main result is a set of equations
between the seepage velocities that together with a constitutive relation
between the average seepage velocity and the driving forces lead to a closed 
set of equations.            
  
We consider only one-dimensional flow in this paper, deferring to later the
generalization to three dimensions. The fluids are assumed to be incompressible.

In section \ref{system} we describe the porous medium system we consider. We
review the key concepts that will be used in the subsequent discussion. In
particular, we discuss the relation between average seepage velocity and the
seepage velocities of each of the two fluids. 

In section \ref{scaling}, we introduce our thermodynamic considerations.  We focus
on the observation that the total volumetric flow rate is an 
Euler homogeneous function of the first order.  
This allows us to define two thermodynamic velocities
as derivatives of the total volumetric flow.  We then go on in section
\ref{saturation} to deriving several equations between the thermodynamic velocities, one
of which is closely related to the Gibbs--Duhem equation in thermodynamics.  In section
\ref{seepage}, we point out that the seepage velocity of each fluid is generally
{\it not\/} equal to the corresponding thermodynamic velocity.  This is due to
the constraints that the geometry of the porous medium puts on how the immiscible fluids
arrange themselves.  In ordinary thermodynamics, such constraints are not present and 
questions of this type do not arise.  We relate the thermodynamic and seepage velocities
through the introduction of a {\it co-moving velocity\/} function.  This
co-moving velocity is a characteristic of the porous medium and depends of the driving
forces only through the velocities and the saturation.
        
We discuss in section \ref{miscibility} what happens when the thermodynamic and seepage 
velocities coincide, when the 
wetting and non-wetting seepage velocities coincide and when the thermodynamic
wetting and non-wetting velocities coincide. Normally, the coincidences appear at
different saturations. However, under certain conditions, the three
coincide.  When this happens, the fluids behave as if they were miscible.   
   
In section \ref{non-eq} we write down the full set of equations
to describe immiscible two-phase flow in porous media.  These equations are
valid on a large scale where the porous medium may be seen as a continuum. 

In section \ref{2examples}, we analyze four versions of the capillary tube 
model \cite{s53,s74} within the concept of the representative elementary volume (REV) 
developed in the previous sections. This allows us to demonstrate these concepts in detail
and to demonstrate the internal consistency of the theory. 

In section \ref{numerical} we analyze a network model \cite{amhb98} for immiscible two-phase
flow within the framework of our theory.  We calculate the co-moving velocity for
the model using a square and a hexagonal lattice.  The co-moving velocity is to within
the level of the statistical fluctuations equal for the two lattice topologies. We
compare successfully the measured seepage velocities for each fluid component with those
calculated from our theory. Lastly, we use our theory to predict the coincidence of
the three saturations defined in section \ref{miscibility}, and verifying this
prediction numerically.         

Section \ref{conclusion} summarizes the results from the previous sections
together with some further remarks on the difference between our approach
and that of relative permeability. We may nevertheless anticipate here what will be
our conclusions.  Through our theory we have accomplished two things.  The first one
is to construct a closed set of equations for the seepage velocities as a function of 
saturation based solely on thermodynamic principles.  We do {\it not\/} propose relations 
between seepage velocities and driving forces such as pressure gradient here.  Our discussion 
is solely based on relations between the seepage velocities.  The second accomplishment is to 
pave the way for further thermodynamic analysis by identifying the proper thermodynamic
variables that relate to the flow rates.     

\section{Defining the system}
\label{system}

The aim of this paper is to derive a set of equations on the continuum level
where differentiation make sense. We define a \textit{representative elementary 
volume\/} --- REV --- as a block of porous material with no internal 
structure filled with two immiscible and incompressible fluids: it is  
described by a small set of parameters which we will now proceed to define.  

\begin{figure*}[tbp]
\includegraphics[width=1.00\textwidth,clip]{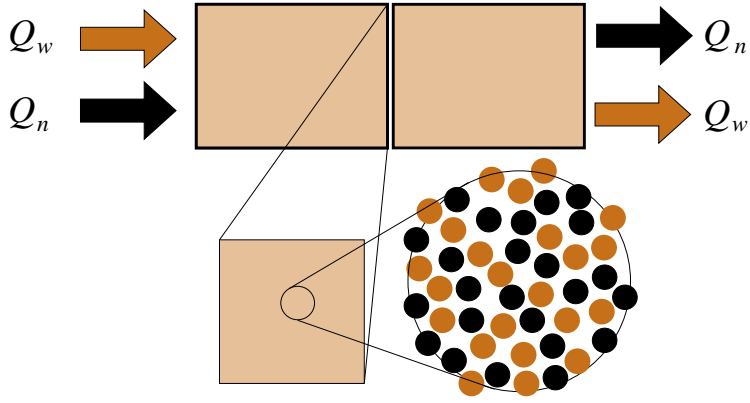}
\caption{In the upper part of the figure, we see the REV from the side. There 
is a flow $Q=Q_w+Q_n$ across it. An imaginary cut is made through the REV 
in the direction orthogonal to the flow. In the lower left corner, the surface 
of the imaginary cut is illustrated. A magnification of the surface of the cut is
shown in the lower right corner. The pore structure is illustrated as brown
and black circles. The pores that are brown, are filled with wetting fluid
and the pores that are black, are filled with non-wetting fluid. The wetting
fluid-filled pores form in total an area $A_w$ and the non-wetting
fluid-filled pores form in total an area $A_n$. The total pore area of the
imaginary cut in the lower left corner is $A_p=A_w+A_n=A\phi$.}
\label{fig1}
\end{figure*}

We illustrate the REV in Fig.\ \ref{fig1}. It is a block of homogeneous porous
material of length $L$ and area $A$. We seal off the surfaces that are
parallel to the $L$-direction (the flow direction). The two remaining
surfaces, each with area $A$, are kept open and act as inlet and outlet for
the fluids that are injected and extracted from the REV. The porosity is 
\begin{equation}
\label{eqn0}
\phi \equiv \frac{V_p}{V}=\frac{V_p}{AL}\;,
\end{equation}
where $V_{p}$ is the pore volume and $V=AL$ the total volume of the REV. 
Due to the homogeneity of the porous
medium, any cross section orthogonal to the axis along the $L$-direction
(named the $x$ axis for later) will reveal a \textit{pore area\/} that
fluctuates around the value 
\begin{equation}
\label{eqn0.1}
A_{p}=\frac{V_p}{L}=\phi A\;,
\end{equation} 
while the solid matrix area
fluctuates around 
\begin{equation}
\label{eqn0.2}
A_s\equiv A-A_{p}=(1-\phi )A\;. 
\end{equation}
The homogeneity assumption consists in the fluctuations being so small that they 
can be ignored.

There is a time averaged volumetric flow rate $Q$ through the REV (see
figure \ref{fig1}). The volumetric
flow rate consists of two components, $Q_{w}$ and $Q_{n}$, which are the
volumetric flow rates of the more wetting ($w$ for 
``wetting") and the less wetting ($n$ for ``non-wetting")
fluids with respect to the porous medium. We have 
\begin{equation}
\label{eqn1}
Q=Q_{w}+Q_{n}\;.  
\end{equation}
All flows have as frame of reference the non-deformable solid matrix.

In the porous medium, there is a volume $V_w$ of \textit{incompressible\/}
wetting fluid and a volume $V_n$ of \textit{incompressible\/} non-wetting fluid 
so that $V_p=V_w+V_n$. We define the wetting and non-wetting saturations 
$S_w=V_w/V_p$ and $S_n=V_n/V_p$ so that 
\begin{equation}  
\label{eqn2}
S_w+S_n=1\;.
\end{equation}

We define the wetting and non-wetting pore areas $A_{w}$ and $A_{n}$ as the
parts of the pore area $A_{p}$ which are filled with the wetting or the
non-wetting liquids respectively. As the porous medium is homogeneous, we
will find the same averages $A_{w}$ and $A_{n}$ in any cross section through
the porous medium orthogonal to the flow direction. This is
illustrated in Fig.\ \ref{fig1}. We have that $%
A_{w}/A_{p}=(A_{w}L)/(A_{p}L)=V_{w}/V_{p}=S_{w}$ so that 
\begin{equation}
\label{eqn3}
A_{w}=S_{w}A_{p}\;.  
\end{equation}
Likewise,
\begin{equation}
\label{eqn4}
A_{n}=S_{n}A_{p}=\left( 1-S_{w}\right) A_{p}\;.  
\end{equation}
where 
\begin{equation}
\label{eqn5}
A_{p}=A_{w}+A_{n}\;.  
\end{equation}

We define the seepage velocities for the two immiscible fluids, $v_w$ and 
$v_n$ as 
\begin{equation}  
\label{eqn5.1}
v_w=\frac{Q_w}{A_w}\;,
\end{equation}
and 
\begin{equation}  
\label{eqn5.2}
v_n=\frac{Q_n}{A_n}\;.
\end{equation}
Hence, equation (\ref{eqn1}) may be written 
\begin{equation}  
\label{eqn14.9}
Q=A_wv_w +A_nv_n\;.
\end{equation}
We finally define a seepage velocity associated with the total flow
rate $Q$ as 
\begin{equation}  
\label{eqn5.3}
v=\frac{Q}{A_p}\;.
\end{equation}
By using equations (\ref{eqn3}) to (\ref{eqn5}) and (\ref{eqn5.1}), 
(\ref{eqn5.2}) and (\ref{eqn5.3}) we transform (\ref{eqn14.9}) into 
\begin{equation}  
\label{eqn5.4}
v=S_w v_w+S_nv_n\;.
\end{equation}
All the variables in this equation can be measured experimentally.

The extensive variables describing the REV, among them the wetting and 
non-wetting pore areas $A_w$ and $A_n$, and the wetting and non-wetting
volumetric flow rates $Q_w$ and $Q_n$, are averages over the REV and their
corresponding intensive variables $S_w$, $S_n$, $v_w$ and $v_n$ are 
therefore by definition a property of the REV.   

This definition of a REV is the same as 
that used by Gray and Miller \cite{gm05} in their thermodynamically 
constrained averaging theory and in the earlier literature on thermodynamics 
of multi-phase flow, see e.g.\ \cite{hg90}. The REV can be regarded as 
homogeneous only on the REV scale.  With one driving force the variables on this scale 
can be obtained, knowing the micro-scale ensemble distribution \cite{sbksvh17,sshbkv17}.  
This averaging procedure must keep invariant the entropy production, a necessary 
condition listed already in the 1990ies \cite{hg90,gm05}.

\section{The volumetric flow rate $Q$ is an Euler homogeneous function of 
order one}
\label{scaling}

\begin{figure*}[tbp]
\includegraphics[width=1.00\textwidth,clip]{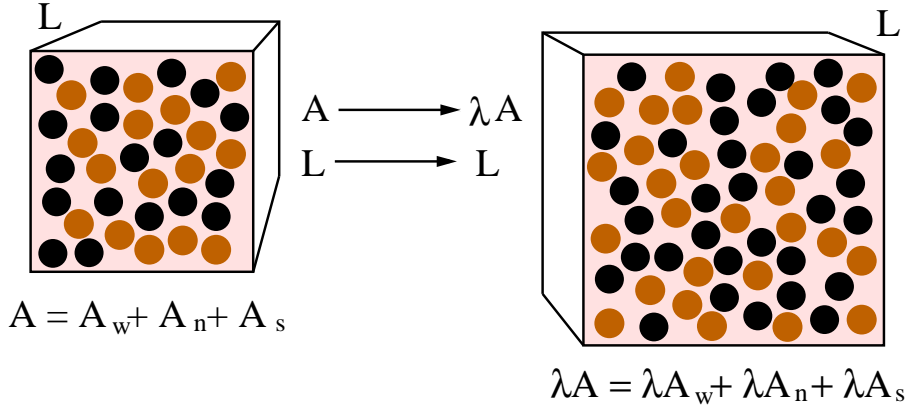} 
\caption{A representative elementary volume (REV) is shown to the left.  The
flow direction is orthogonal to the figure plane. The REV has a length $L$
in the flow direction and an area $A$ orthogonal to the flow direction.  That
area may be split in three: $A_s$ which is the area of the solid matrix (colored
pink), $A_w$ the area covered by pores filled with wetting fluid (colored 
brown), and $A_n$ the area covered by pores filled with non-wetting fluid
(colored black). To the right we show a different REV with an area 
$\lambda A$ orthogonal to the flow direction.  The size of the pores in this 
REV is the same as in the one to the left as we assume them to be made of the 
same porous material.  The solid matrix, wetting and non-wetting areas in this 
REV are then $\lambda A_s$, $\lambda A_w$ and $\lambda A_n$ respectively. The 
length of the REV is the same as the one to the left, $L$.  Keeping the local
(intensive) parameters of the flow fixed when rescaling the areas, leads to a
change in the volumetric flow rate from $Q$ (left) to $\lambda Q$ (right). 
This illustrates the contents of equation (\ref{eqn10}).}
\label{fig2}
\end{figure*}

The volumetric flow rate $Q$ is a {\it homogeneous function of order one\/} 
of the areas $A_{w}$ and $A_{n}$ defined in Section \ref{system}.  We now 
proceed to demonstrate this. 
Suppose we scale the three areas $A_{w}\rightarrow \lambda A_{w}$, 
$A_{n}\rightarrow \lambda A_{n}$ and $A_s\rightarrow \lambda A_s$,
keeping in mind that
\begin{equation}
\label{eqn9.5}
A_s=\left(\frac{1}{\phi}-1\right)\ A_p=\left(\frac{1}{\phi}-1\right)\left(A_w+A_n\right)\;,
\end{equation}  
which we get from equations (\ref{eqn0.1}) and (\ref{eqn0.2}).  This 
corresponds to enlarging the area $A$ of the REV to $\lambda A$ as illustrated 
in figure \ref{fig2}.  This transformation does not change the porosity $\phi$.
Due to equation (\ref{eqn9.5}) with $\phi$ kept fixed, the only two independent 
variables in the problem are then $A_w$ and $A_n$, and as a consequence the volumetric
flow rate depends on these two variables and not on $A_s$, $Q=Q(A_w,A_n)$.  
The scaling $A_w\to \lambda A_w$ and
$A_n\to\lambda A_n$ affects the volumetric flow rate as follows:
\begin{equation}
\label{eqn10}
Q(\lambda A_{w},\lambda A_{n})=\lambda Q(A_{w},A_{n})\;.
\end{equation}
As long as the porous material does not have a fractal structure, this scaling 
property is essentially self evident as explained in Fig.\ \ref{fig2}. 

We take the derivative with respect to $\lambda $ on both sides of 
(\ref{eqn10}) and set $\lambda =1$. This gives\footnote{We clarify the
physical meaning of these derivatives through a 
number of examples in section \ref{2examples}.} 
\begin{equation}
\label{eqn11}
Q(A_{w},A_{n})=A_{w}\left( \frac{\partial Q}{\partial A_{w}}\right)
_{A_{n}}+A_{n}\left( \frac{\partial Q}{\partial A_{n}}\right)
_{A_{w}}\;.
\end{equation}
Equation (\ref{eqn11}) is essentially
the Euler theorem for homogeneous functions of order one. By dividing this
equation by $A_{p}$, we have 
\begin{equation}
\label{eqn10-1}
v=S_{w}\left( \frac{\partial Q}{\partial A_w}\right)
_{A_n}+S_{n}\left( \frac{\partial Q}{\partial A_n}\right)_{A_w}\;,  
\end{equation}
where we have used equations (\ref{eqn3}) and (\ref{eqn4}).

The two partial derivatives in equation (\ref{eqn10-1}) have the units of
velocity.  Hence, we define two {\it thermodynamic velocities\/} $\hat{v}_w$ and $\hat{v}_n$ as
\begin{equation}
\label{eqn10-2}
\hat{v}_{w}=\left( \frac{\partial Q}{\partial A_{w}}\right)_{A_{n}}\;,  
\end{equation}
and
\begin{equation}
\label{eqn10-3}
\hat{v}_{n}=\left( \frac{\partial Q}{\partial A_{n}}\right)_{A_{w}}\;.  
\end{equation}
Equation (\ref{eqn10-1}) then becomes
\begin{equation}
\label{eqn10-5}
v=S_w \hat{v}_w+S_n\hat{v}_n\;.
\end{equation}

The fact that the  volumetric flow rate $Q$
is a homogeneous function of order one in the variables $A_w$ and
$A_n$ implies that the thermodynamic 
velocities defined in equations (\ref{eqn10-2})
and (\ref{eqn10-3}) must be homogeneous of the zeroth order so that
have  
\begin{eqnarray}
&&\hat{v}_{w}(\lambda A_{w},\lambda A_{n})=\hat{v}_{w}(A_{w},A_{n})\;,\nonumber\label{eqn14a} \\
&&\hat{v}_{n}(\lambda A_{w},\lambda A_{n})=\hat{v}_{n}(A_{w},A_{n})\;,  \nonumber \\
&&v(\lambda A_{w},\lambda A_{n})=v(A_{w},A_{n})\;.
\end{eqnarray} 
This implies that they must depend on the areas $A_{w}$ and $A_{n}$ through their ratio 
$A_{w}/A_{n}=S_{w}/S_{n}=S_{w}/(1-S_{w})$, where we have used equations (\ref%
{eqn3}) --- (\ref{eqn5}) and we may write 
\begin{eqnarray}
&&\hat{v}_{w}=\hat{v}_{w}(S_{w})\;,  \nonumber  \label{eqn14b} \\
&&\hat{v}_{n}=\hat{v}_{n}(S_{w})\;,  \nonumber \\
&&v=v(S_{w})\;.
\end{eqnarray}

\section{Consequences of the Euler theorem: new equations}
\label{saturation}

We now change the variables from $(A_{w},A_{n})$ to $(S_{w},A_{p})$ where 
$A_{w}=S_{w}A_{p}$ and $A_{n}=(1-S_{w})A_{p}$.  We calculate 
\begin{eqnarray}
\label{eqn18.0}
\left(\frac{\partial Q}{\partial S_{w}}\right) _{A_{p}}&=&
\left(\frac{\partial Q}{\partial A_{w}}\right) _{A_{n}}
\left(\frac{\partial A_{w}}{\partial S_{w}}\right) _{A_{p}}+
\left(\frac{\partial Q}{\partial A_{n}}\right)_{A_{w}}
\left(\frac{\partial A_{n}}{\partial S_{w}}\right) _{A_{p}}
\nonumber\\
&=&A_{p}\left[ \hat{v}_{w}-\hat{v}_{n}\right] \;.\\  
\nonumber
\end{eqnarray}
We divide by the area $A_{p}$ and find 
\begin{equation}
\label{eqn18}
\frac{dv}{dS_{w}}=\hat{v}_{w}-\hat{v}_{n}\;,  
\end{equation}
replacing the partial derivative by a total derivative in accordance with 
equation (\ref{eqn14b}).

We then proceed by taking the derivative of equation 
({\ref{eqn10-5}) with respect to $S_w$, 
\begin{equation}  
\label{eqn19}
\frac{dv}{dS_w}=\frac{d}{dS_w}\left[S_w \hat{v}_w+(1-S_w)\hat{v}_n\right] 
=\hat{v}_w-\hat{v}_n+S_w 
\frac{d\hat{v}_w}{dS_w}+(1-S_w)\frac{d\hat{v}_n}{dS_w}\;.
\end{equation}
Combining this equation with equation (\ref{eqn18}), we find 
\begin{equation}  
\label{eqn20}
S_w \frac{d\hat{v}_w}{dS_w}+(1-S_w)\frac{d\hat{v}_n}{dS_w}=0\;.
\end{equation}
This equation is a {\it Gibbs--Duhem\/}-like equation, here
relating the thermodynamic velocities $\hat{v}_w$ and $\hat{v}_n$.

We note that equations (\ref{eqn10-5}), (\ref{eqn18}) and (\ref{eqn20})
are interrelated in that any pair selected from the three equations will contain the third.

The thermodynamic velocities (\ref{eqn10-2}) and (\ref{eqn10-3}) may be expressed in terms of
the variables $A_p$ and $S_w$,
\begin{eqnarray}
\label{eqn10-2-e}
\hat{v}_{w}&=&\left(\frac{\partial Q}{\partial A_{w}}\right)_{A_{n}}
=\left(\frac{\partial A_p v}{\partial A_p}\right)_{S_w}
\left(\frac{\partial A_p}{\partial A_w}\right)_{A_n}
+\left(\frac{\partial A_p v}{\partial S_w}\right)_{A_p}
\left(\frac{\partial S_w}{\partial A_w}\right)_{A_n}\nonumber\\
&=&v+S_n\frac{dv}{dS_w}\;,\nonumber\\
\end{eqnarray}
where we have used that
\begin{equation}
\label{eqn10-2-f}
\left(\frac{\partial S_w}{\partial A_w}\right)_{A_n}=\frac{1}{A_p}-\frac{A_w}{A_p^2}=\frac{S_n}{A_p}\;.
\end{equation}
Likewise, we find
\begin{equation}
\label{eqn10-3-e}
\hat{v}_{n}=\left(\frac{\partial Q}{\partial A_{n}}\right)_{A_{w}}=v-S_w\frac{dv}{dS_w}\;.  
\end{equation}

We introduce a function $g=g(S_w)$ defined as
\begin{equation}\label{eqn20-2}
g(S_w)=-\ \frac{1}{S_w}\ \frac{d\hat{v}_n}{dS_w}\;.
\end{equation}
Combining this equation with equation (\ref{eqn20}) gives 
\begin{equation}
\label{eqn20-3}
\frac{d\hat{v}_w}{dS_w}= S_n g(S_w)\;.
\end{equation}
We now take the derivative of equation (\ref{eqn18}) with respect to $S_w$ 
and combine with the two previous equations
(\ref{eqn20-2}) and (\ref{eqn20-3}) to find
\begin{equation}
\label{eqn20-4}
\frac{d^2v}{dS_w^2}=\frac{d\hat{v}_w}{dS_w}-\frac{d\hat{v}_n}{dS_w}=g(S_w)\;,
\end{equation}
where we have used equation (\ref{eqn2}).  Hence, we have finally
\begin{equation}
\label{eqn20-5}
\frac{d\hat{v}_w}{dS_w}=S_n\ \frac{d^2v}{dS_w^2}\;,
\end{equation}
and
\begin{equation}
\label{eqn20-6}
\frac{d\hat{v}_n}{dS_w}=-S_w\ \frac{d^2v}{dS_w^2}\;.
\end{equation}

\subsection{New equations in terms of the seepage velocities}
\label{seepage}

Equation (\ref{eqn10-5}) has the same form as equation (\ref{eqn5.4}),  
\begin{equation}
\label{eqn55.5}
v=S_w v_w+S_n v_n=S_w\hat{v}_w+S_n \hat{v}_n\;.
\end{equation}    
This does {\it not\/} imply that $\hat{v}_w=v_w$ and 
$\hat{v}_n=v_n$.  The most general relations we can write down between
$\hat{v}_w$ and $v_w$, and $\hat{v}_n$ and $v_n$ and still
fulfill (\ref{eqn55.5}) are 
\begin{equation}
\label{eqn10-5.5}
\hat{v}_{w}=v_{w}+S_n v_m(S_w)\;,  
\end{equation}
and 
\begin{equation}
\label{eqn10-5.6}
\hat{v}_{n}=v_{n}-S_w v_m(S_w)\;,  
\end{equation} 
where $v_m$ is a function of $S_w$ with the units of velocity.
We name this function {\it the co-moving velocity.\/} We may interpret 
the physical contents of the second equality in (\ref{eqn55.5}) in
the following way: The thermodynamic velocities $\hat{v}_w$ and $\hat{v}_n$ 
differ from the $v_w$ and $v_n$ as each fluid carries along
some of the other fluid --- hence ``co-moving." To preserve the volume, 
these co-moving contributions are related in such a way that $A_n S_w v_m$ is 
the co-moving volume flow of the non-wetting fluid with the wetting fluid 
and $A_w S_n v_m $ is the counter-moving volume flow of the wetting fluid with 
the non-wetting fluid. 

By using equations (\ref{eqn10-5.5}) and (\ref{eqn10-5.6}), we find that the seepage 
velocities (\ref{eqn5.1}) and (\ref{eqn5.2}) also are homogeneous of the zeroth 
order so that 
\begin{eqnarray}
&&{v}_{w}={v}_{w}(S_{w})\;,  \nonumber  \label{eqn14c} \\
&&{v}_{n}={v}_{n}(S_{w})\;.
\end{eqnarray}

Equations (\ref{eqn10-5}), (\ref{eqn18}) and (\ref{eqn20}) are in terms of the 
thermodynamic velocities defined by the area derivatives of the volumetric flow, 
(\ref{eqn10-2}) and (\ref{eqn10-3}).  We now express them in terms of the seepage
velocities $v_w$ and $v_n$ defined in equations (\ref{eqn5.1}) and (\ref{eqn5.2}).
We do this by invoking the transformations (\ref{eqn10-5.5}) and (\ref{eqn10-5.6}).
Equation (\ref{eqn18}) then becomes
\begin{equation}
\label{eqn18-1}
\frac{dv}{dS_{w}}={v}_{w}-{v}_{n}+v_m\;,  
\end{equation}
where we have used equations (\ref{eqn10-5.5}) and (\ref{eqn10-5.6}) and 
$v_m$ is the co-moving velocity.

Expressing (\ref{eqn20}) in terms of the seepage velocities (\ref{eqn5.1}) and 
(\ref{eqn5.2}), we find
\begin{equation}  
\label{eqn20-1}
S_w \frac{d{v}_w}{dS_w}+(1-S_w)\frac{d{v}_n}{dS_w}=v_m\;.
\end{equation}

The three equations (\ref{eqn5.4}), (\ref{eqn18-1}) and (\ref{eqn20-1}) are
dependent: combine any two of them and the third will follow.  

By a derivation similar to that which led to equations (\ref{eqn20-5}) 
and (\ref{eqn20-6}), we find
\begin{equation}
\label{eqn20-7}
\frac{dv_w}{dS_w}=v_m+S_n\left[\frac{d^2v}{dS_w^2}-\frac{dv_m}{dS_w}\right]\;,
\end{equation}
and
\begin{equation}
\label{eqn20-8}
\frac{dv_n}{dS_w}=v_m-S_w\left[\frac{d^2v}{dS_w^2}-\frac{dv_m}{dS_w}\right]\;.
\end{equation}

\section{Cross points}
\label{miscibility}

We may define three wetting saturations of special interest, $S_w^A$, $S_w^B$ and $S_w^C$, that 
have the following properties:
\begin{itemize}
  \item{\bf A} When $S_w=S_w^A$ we have $v_w=\hat{v}_w$ and $v_n=\hat{v}_n$.  By 
  equations (\ref{eqn10-5.5}) and (\ref{eqn10-5.6}), we have that 
  $v_m=0$ for this wetting saturation.
  We furthermore have that $v_m\le 0$ for $S_w\le S_w^A$ and $v_m\ge 0$ for $S_w\ge S_w^A$.\\

  \item{\bf B} When $S_w=S_w^B$, we have that $\hat{v}_w=\hat{v}_n$.  From 
  equation (\ref{eqn55.5}) we find $\hat{v}_w=\hat{v}_n=v$.  Equations
  (\ref{eqn10-2-e}) and (\ref{eqn10-3-e}) make $dv/dS_w=0$ equivalent to the
  equality between the three velocities at this wetting saturation.  
  Furthermore, at $dv/dS_w=0$, $v$ has its
  minimum value.  We have that $dv/dS_w\le 0$ for $S_w\le S_w^B$ and $dv/dS_w\ge 0$ for
   $S_w\ge S_w^B$.\\ 

  \item{\bf C} When $S_w=S_w^C$, we have that $v_w=v_n$. This implies that 
  $v_w=v_n=v$ from equation (\ref{eqn5.4}) for this wetting saturation. 
  By equations (\ref{eqn11-5.5})
  and (\ref{eqn11-5.6}) we have that this is equivalent to $v_m=dv/dS_w$.\\  
\end{itemize}    

By considering the relations between the velocity variables derived so far, 
we may show that we either have
\begin{equation}
\label{mp1}
S_w^A\le S_w^B\le S_w^C\;,
\end{equation}
or
\begin{equation}
\label{mp2}
S_w^C\le S_w^B\le S_w^A\;.
\end{equation}

It is straight forward to demonstrate that if either two of the three
wetting saturations $S_w^A$, $S_w^B$ or $S_w^C$, coincide, then the 
third must also have the same value.  If this is the case, we have $v_w=v_n$ and 
$dv/dS_w=0$, so that the two immiscible fluids behave as if they were miscible 
\cite{sgvwhf17}.  We will use our theory to identify the wetting saturation
at which this coincidence occurs 
for the network model we study in section \ref{numerical} and then verify numerically 
that this is indeed so.    

\section{A closed set of equations}
\label{non-eq}

We now consider the porous medium on scales larger than the
REV scale.  The properties of the porous medium may at these scales vary in space.
We consider here only flow in
the $x$ direction, deferring the generalization to higher dimensions
to later.  Let $x$ be a point somewhere in 
this porous medium. Hence, all the variables in the following will be
functions of $x$. 
  
Equations (\ref{eqn5.4}) and (\ref{eqn20-1}), 
\begin{eqnarray}
v&=&S_w v_w+S_n v_n\;,\nonumber\\
v_m&=&S_w v_w'+S_n v_n'\;,\nonumber\\
\nonumber
\end{eqnarray}
where $v_w'=dv_w/dS_w$ and $v_n'=dv_n/dS_w$, can be seen as a transformation
from the velocity pair $(v_w,v_n)$ to the pair $(v,v_m)$.
We invert the velocity transformation, 
$(v,v_m)\to (v_w,v_n)$, to find
\begin{equation}
\label{eqn11-5.5}
v_w=v +S_n (v'-v_m)\;,
\end{equation}
and 
\begin{equation}
\label{eqn11-5.6}
v_n=v -S_w (v'-v_m)\;,
\end{equation}
where $v'=dv/dS_w$.   

We now turn to the equations controlling the time evolution of the flow.
The transport equation for $S_w$ is  
\begin{equation}  
\label{conc-1}
\frac{\partial S_w}{\partial t}+\frac{\partial S_w v_w}{\partial x}=0\;.
\end{equation}
Likewise, we have for $S_n$
\begin{equation}  
\label{conc-1.10}
\frac{\partial S_n}{\partial t}+\frac{\partial S_n v_n}{\partial x}=0\;.
\end{equation}
Summing these two equations and using equations (\ref{eqn2}) and 
(\ref{eqn5.4}), we find the incompressibility condition
\begin{equation}
\label{conc-1.1}
\frac{\partial v}{\partial x}=0\;.
\end{equation}

Equations (\ref{eqn5.4}), (\ref{eqn20-1}), (\ref{conc-1}) and (\ref{conc-1.1}) 
constitute four equations.  There are five variables: $v$, $v_m$, $v_w$, $v_n$ and $S_w$.
Seemingly, we are one equation short of a closed set. 

In order to close the equation set, we need an equation for the co-moving velocity 
$v_m$.  This is a {\it constitutive equation\/} that characterizes how the immiscible 
fluids affect each other due to the constraints imposed by the porous matrix, 
\begin{equation}
\label{conc-8}
v_m=v_m(S_w,v,v')\;.
\end{equation}
Specifying $v_m$ will specify the form of the differential equation 
(\ref{eqn20-1}).  We will work out $v_m$ analytically for four systems in 
section \ref{2examples} and map it for a range of parameters in a network 
model in section \ref{numerical}.      

Note that it is not enough to specify $S_w$ and $v$ to determine $v_m$. It is
also a function of $dv/dS_w$, since $dv/dS_w$ depends on how the external 
parameters are controlled as $S_w$ is changed.  For example, if the total
volumetric flow rate $Q=A_p v$ is held constant when $S_w$ is changed,
then $dv/dS_w=0$ for all values of $S_w$, and the system traces the curve
$(S_w,v,v'=0)$ in $(S_w,v,v')$ space. If, on the other hand, the
driving forces are held constant when $S_w$ is changed, then $v$ and $dv/dS_w$ will 
follow some non-trivial curve in $(S_w,v,v')$ space, see section \ref{numerical}.  
A third example is to control the wetting and the non-wetting volumetric
flow rates $Q_w$ and $Q_n$, making $S_w$ and $v$ and $dv/dS_w$ dependent variables, 
following curves in $(S_w,v,v')$ space depending on
how $Q_w$ and $Q_n$ are changed.

The driving forces enter the discussion through the constitutive equation  
\begin{equation}  
\label{conc-7}
v=v(S_w,f)\;,
\end{equation}  
where $f$ represents the driving forces, e.g., $f=-\partial P/\partial x$, where $dP/dx$ is 
the pressure gradient.

The equation set (\ref{eqn5.4}), (\ref{eqn20-1}), (\ref{conc-1}) 
and (\ref{conc-1.1}) together with the constitutive equation for
$v_m$ is {\it independent\/} of the driving forces.  They are valid for
any constitutive equations (\ref{conc-7}) that may be proposed.
In particular, non-linear constitutive equations such as
those recently proposed which suggest that two immiscible fluids behave as if 
they were a single Bingham plastic 
\cite{tkrlmtf09,tlkrfm09,rcs11,sh12,sbdkpbtsch16} fits well into this
framework. 

This is in stark contrast to e.g.\ the
relative permeability framework that assumes a particular constitutive
equation for $v_w$ and $v_n$ containing the two relative
permeabilities $k_{r,w}(S_w)$ and $k_{r,n}(S_w)$ in addition to the
capillary pressure function $P_c(S_w)$.  The equation set we present 
here is thus much more general than that of the relative permeability
framework.  

\section{Analytically tractable models}
\label{2examples}

We will in this section analyze four variants of the capillary tube 
model \cite{s53,s74}. In each case, we calculate the co-moving velocity
$v_m$ and then use it to demonstrate the consistency of our theory.  We also
use this section to clarify the physical meaning of the area derivatives
introduced in section \ref{scaling}. 

\subsection{Parallel capillaries filled with either fluid}
\label{simplest}

In this simplest case, we envision a bundle of $N$ parallel capillaries, all equal. Each 
tube has a cross-sectional inner (i.e.\ pore) area $a_p$.
$N_w$ of these capillaries are filled with the wetting fluid and $N_n$ 
are filled with the non-wetting fluid, so that $N_w+N_n=N$.   Let us assume that 
the seepage velocity in the capillaries filled with wetting fluid is $v_w$ and the 
seepage velocity in the capillaries filled 
with non-wetting fluid is $v_n$.  The total volumetric flow rate is then
\begin{equation}
\label{sim-1}
Q=N_w a_p v_w + N_n a_p v_n\;.
\end{equation}
The wetting saturation is given by
\begin{equation}\label{sim-2}
S_w=\frac{A_w}{A_p}=\frac{N_w}{N}\;.
\end{equation}
We calculate the derivative of $Q$ with respect to $A_w$ to determine the thermodynamic
velocity
$\hat{v}_w$ defined in equation (\ref{eqn10-2}) together
with equation (\ref{sim-1}),
\begin{equation}
\label{sim-3}
\hat{v}_w=\left(\frac{\partial Q}{\partial A_w}\right)_{A_n}=
\frac{1}{a_p\delta N_w}\ \left[Q(N_w+\delta N_w,N_n)-Q(N_w,N_n)\right]=v_w\;.
\end{equation}
The derivative is thus performed changing the number of capillaries from $N$ to
$N+\delta N$ and letting all the added capillaries contain the wetting fluid
so that $N_w\to N_w+\delta N_w=N_w+\delta N$.  Likewise, we find using equation
(\ref{eqn10-3})
\begin{equation}
\label{sim-4}
\hat{v}_n=\left(\frac{\partial Q}{\partial A_n}\right)_{A_w}=
\frac{1}{a_p\delta N_n}\ \left[Q(N_w,N_n)-Q(N_w,N_n+\delta N_n)\right]=v_n\;.
\end{equation}
From equations (\ref{eqn10-5.5}) and (\ref{eqn10-5.6}) we find that the
co-moving velocity is zero,
\begin{equation}
\label{sim-5}
v_m=0\;
\end{equation}
for this system. Hence, we have that the seepage velocities $v_w$ and $v_n$ are
equal to the thermodynamic velocities $\hat{v}_w$ and $\hat{v}_n$.

\begin{figure*}[tbp]
\includegraphics[width=0.6\textwidth,clip]{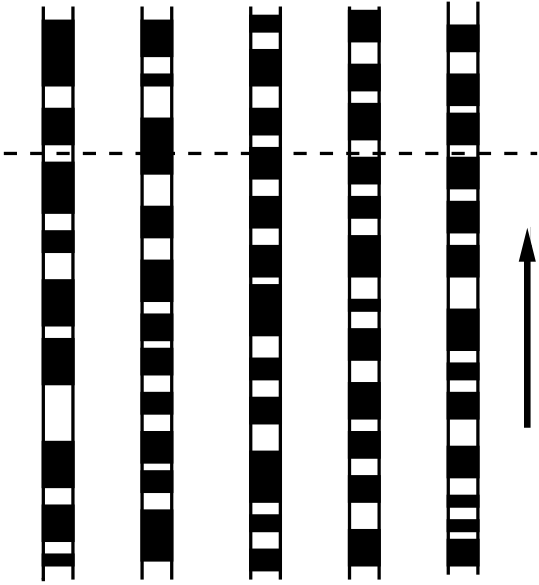}
\caption{The parallel tube model analyzed in section \ref{bubbles}. There are $N=5$ capillaries, 
each filled
with a bubble train of wetting (white) and non-wetting (black) fluid moving in the direction of the 
arrow. The diameter along each tube vary so that the capillary force from each interface vary with its
position.  The varying diameters are not illustrated in the figure; only the average diameter is shown.
This is the same for all of the capillaries.  Each capillary has a length $L$ and an inner area $a_p$.  The
wetting fluid in each tube occupies a volume $L_w a_p$, whereas the non-wetting fluid occupies a 
volume $L_n a_p$. The broken line illustrates the imaginary cut through the capillary tube bundle.}
\label{fig3}
\end{figure*}

\subsection{Parallel capillaries with bubbles}
\label{bubbles}

Suppose that we have $N$ parallel capillaries as shown in Fig.\ \ref{fig3}.  Each tube has a length
$L$ and an average inner area $a_p$. The diameter of each capillary varies along the long axis.  
Each capillary
is filled with a bubble train of wetting or non-wetting fluid.  The capillary forces due to the
interfaces vary as the bubble train moves due to the varying diameter. We furthermore assume that
the wetting fluid, or more precisely, the {\it more wetting fluid\/} 
(in contrast to the non-wetting fluid which is the {\it less wetting fluid\/}) does not form
films along the pore walls so that the fluids do not pass each other, see \cite{shbk13} for details.
The volume of the wetting fluid in each tube is $L_w a_p$ and the volume of the non-wetting 
fluid is $L_n a_p$.  Hence, the saturations are $S_w=L_w/L$ and $S_n=L_n/L$ for each tube.  

Suppose now that the seepage velocity in each tube is $v$ when averaged over time.  
Both the wetting and non-wetting seepage velocities must be equal to the average seepage 
velocity since the bubbles do not pass each other:
\begin{equation}
\label{para10}
v_w=v_n=v\;.
\end{equation}        

We now make an imaginary cut through the capillaries orthogonal to the flow direction 
as shown in Fig.\ \ref{fig3}. There will be a number $N_w$ capillaries where the cut 
passes through the wetting fluid and a number $N_n$ capillaries where the cut passes 
through the non-wetting fluid. Averaging over time, we must have
\begin{equation}
\label{para11}
\frac{\langle N_w\rangle}{N} = \frac{L_w}{L}=S_w
\end{equation}
and
\begin{equation}
\label{para12}
\frac{\langle N_n\rangle}{N} = \frac{L_n}{L}=S_n\;.
\end{equation}
Hence, the wetting and non-wetting areas defined in section \ref{system} are
\begin{equation}
\label{para13}
A_w=\langle N_w\rangle a_p
\end{equation}
and
\begin{equation}
\label{para13-2}
A_n=\langle N_n\rangle\ a_p;.
\end{equation}
We also have that 
\begin{equation}
\label{para13-1}
A_p=N a_p\;.
\end{equation}

We will now calculate the derivative (\ref{eqn10-2}) defining the thermodynamic 
velocity $\hat{v}_w$. 
We do this by changing the pore area $A_p\to A_p+\delta A_p=Na_p+a_p \delta N$.
We wish to change $A_w$ while keeping $A_n$ fixed. This can only be done by adjusting 
$S_w$ while changing $\delta N$.  This leads to the two equations 
\begin{equation}
\label{para14}
\delta \langle N_w\rangle =\delta[N S_w]=N\delta S_w+S_w\delta N=\delta N\;,
\end{equation}
and
\begin{equation}
\label{para15}
\delta \langle N_n\rangle =\delta[N (1-S_w)]=-N\delta S_w+(1-S_w)\delta N=0\;.
\end{equation}

We solve either (\ref{para14}) or (\ref{para15}) for $\delta S_w$ (they 
contain the same information), finding
\begin{equation}
\label{para16}
\delta S_w=\frac{(1-S_w)}{N}\ \delta N\;.
\end{equation}

Hence, we have
\begin{eqnarray}
\label{para17}
\hat{v}_w&=&\left(\frac{\partial Q}{\partial A_w}\right)_{A_n}
=\frac{1}{\delta A_p}\ \left[(A_p+\delta A_p) v(S_w+\delta S_w)-A_pv(S_w)\right]\nonumber\\
&=&\frac{1}{\delta N}\ \left[(N+\delta N) v(S_w+\delta S_w)-Nv(S_w)\right]\nonumber\\
&=&\frac{1}{\delta N}\ \left[(N+\delta N) [v(S_w)+(dv/dS_w)\delta S_w]-Nv(S_w)\right]\nonumber\\
&=&\frac{1}{\delta N}\ \left[(N+\delta N) [v(S_w)+(1-S_w)(dv/dS_w)(\delta N/N)]-Nv(S_w)\right]\nonumber\\
&=&v(S_w)+S_n\ \frac{dv(S_w)}{dS_w}\;.\nonumber\\
\end{eqnarray}

We recognize that this equation is equation (\ref{eqn10-2-e}).  Hence, we could have taken this 
equation as the starting point of discussing this model. However, doing the derivative explicitly
demonstrates its operational meaning. 

Rather than performing a similar calculation for $\hat{v}_n$ as in (\ref{para17}) for
$\hat{v}_w$, we use equation (\ref{eqn10-3-e}) and have   
\begin{equation}
\label{para21}
\hat{v}_n=\left(\frac{\partial Q}{\partial A_n}\right)_{A_w}=v(S_w)-S_w\ \frac{dv(S_w)}{dS_W}\;.
\end{equation}

We now combine equations (\ref{para10}), (\ref{para17}) and (\ref{para21}) with 
(\ref{eqn10-5.5}) and (\ref{eqn10-5.6}) and read off
\begin{equation}
\label{para22}
v_m=\frac{dv}{dS_w}\;.
\end{equation}

We see that the expression for $v_m$ is independent of the constitutive equation that 
relates flow rate to the driving forces.  Rather, it expresses the relation between
the thermodynamic velocities $\hat{v}_w$ and $\hat{v}_n$, and the seepage velocities, reflecting the
co-moving --- or rather the lack of it --- in the capillaries.     

Combining equation (\ref{para22}) for $v_m$ with equations (\ref{eqn10-5.5}) and (\ref{eqn10-5.6})
gives equation (\ref{para10}) as it must.

This result gives us an opportunity to clarify the physical meaning of the co-moving 
velocity $v_m$.  We have assumed that the bubbles cannot pass each other as 
shown in Fig.\ \ref{fig3}. 
Equation (\ref{para22}) gives upon substitution in equations (\ref{eqn10-5.5}) and 
(\ref{eqn10-5.6}), 
\begin{equation}
\label{last10}
v_w=\hat{v}_w-S_n\frac{dv}{dS_w}\;,
\end{equation}
and
\begin{equation}
\label{last11}
v_n=\hat{v}_n+S_w\frac{dv}{dS_w}\;.
\end{equation}
We see that these two equations compensate exactly for the two
equations (\ref{eqn10-2-e}) and (\ref{eqn10-3-e}),
\begin{eqnarray}
\hat{v}_w&=&v+S_n\frac{dv}{dS_w}\;,\nonumber\\
\hat{v}_n&=&v-S_w\frac{dv}{dS_w}\;,\nonumber\\
\nonumber
\end{eqnarray}
that give the thermodynamic velocities resulting in $v_w=v_n=v$.

\subsection{Parallel capillaries with a subset of smaller ones}
\label{cfb}

We now turn to the third example.  There are $N$ parallel
capillaries of length $L$.  A fraction of these capillaries has an inner area
$a_s$, whereas the rest has an inner area $a_l$.  The total pore area carried
by the small capillaries is $A_s$ and the total pore area carried by the
larger capillaries is $A_l$ so that
\begin{equation}
\label{para23}
A_p=A_s+A_l\;,
\end{equation}
We define the fraction of small capillaries to be $S_{w,i}$, so that
\begin{equation}
\label{para24}
A_s=S_{w,i} A_p\;.
\end{equation}          

We now assume that the small capillaries are so narrow that only the wetting 
fluid enters them.  These pores constitute the irreducible wetting-fluid 
contents of the model in that we cannot go below this saturation. However,
they still contribute to the flow.   Hence, at a saturation $S_w$ which we assume to be larger
than or equal to $S_{w,i}$ which is then the {\it irreducible wetting fluid saturation\/}, 
we have that the wetting pore area is 
\begin{equation}
\label{para25}
A_w=S_{w,i} A_p+(S_w-S_{w,i})A_p=A_s+A_{lw}\;,
\end{equation}
where we have defined $A_{lw}=(S_w-S_{w,i})A_p$.

We assume there is a seepage
velocity $v_{sw}$ in the small capillaries, a seepage velocity $v_{lw}$ in the
larger capillaries filled with wetting fluid and a seepage velocity $v_n$ in the larger
filled with non-wetting fluid.  The velocities $v_{sw}$, $v_{lw}$ and $v_n$ 
are independent of each other. The total volumetric flow rate is then given by
\begin{equation}
\label{para26}
Q=A_s v_{sw}+A_{lw} v_{lw}+A_n v_{n}\;,
\end{equation}
or in term of average seepage velocity
\begin{equation}
\label{para26-1}
v=S_{w,i} v_{sw}+(S_w-S_{w,i}) v_{lw}+S_n v_{n}\;.
\end{equation}

We will now calculate the derivative (\ref{eqn10-2}) defining the
thermodynamic velocity $\hat{v}_w$.  
We could have done this using equation (\ref{eqn10-2-e}). 
However, it is instructive to perform the derivative yet again through area differentials
so that their meaning become clear.
  
In order to calculate the derivative (\ref{eqn10-3}) we change the pore area $A_p\to A_p+\delta A_p$ so that 
$\delta A_w=\delta A_p$ and $\delta A_n=0$. In 
addition, we have that $\delta A_s=S_{w,i} \delta A_p$, so that
\begin{equation}
\label{para27}
\delta A_w=\delta A_s+\delta A_{lw}=S_{w,i} \delta A_p+\delta A_{lw}=\delta A_p\;.  
\end{equation}
Hence, we have 
\begin{equation}
\label{para28}
\delta A_{lw}=(1-S_{w,i})\delta A_p\;.
\end{equation}
We now combine this expression with equation (\ref{para26}) 
for the total volumetric flow $Q$ to find
\begin{eqnarray}
\label{para29}
\hat{v}_w&=&\left(\frac{\partial Q}{\partial A_w}\right)_{A_n}
=\frac{1}{\delta A_p}\ \left[\delta A_s v_{sw} + \delta A_{lw} v_{lw} 
- A_s v_{sw}\right]\nonumber\\
&=&v_{lw}+S_{w,i} (v_{sw}-v_{lw}).\nonumber\\
\end{eqnarray}

We could have found $\hat{v}_n$ in the same way.  However, using (\ref{eqn10-3-e})
combined with (\ref{para26-1}) gives 
\begin{equation}
\label{para30}
\hat{v}_n=v_{n}+S_{w,i}(v_{sw}-v_{lw})\;.
\end{equation}

We now use equation (\ref{eqn10-5.6}) to find
\begin{equation}
\label{para31}
\hat{v}_n-v_n= - v_m S_w = S_{w,i}(v_{sw}-v_{lw})=-\left[\frac{S_{w,i}}{S_w}(v_{lw}-v_{sw})\right] S_w\;,
\end{equation}
so that
\begin{equation}
\label{para32}
v_m=\frac{S_{w,i}}{S_w}\left(v_{lw}-v_{sw}\right)\;.
\end{equation}

As a check, we now use the co-moving velocity found in (\ref{para32}) together with
equation (\ref{eqn10-5.5}) to calculate $v_w$.  We find
\begin{equation}
\label{para33}
v_w=\hat{v}_w-v_m S_n=\frac{1}{S_w}\left[S_{w,i} v_{sw}+(S_w-S_{w,i}) v_{lw}\right]\;,
\end{equation}
which is the expected result.  

\subsection{Large capillaries with bubbles and small capillaries with wetting fluid only}
\label{cfbbubbles}

The fourth example that we consider is a combination of the two previous models discussed in
section \ref{bubbles} and \ref{cfb}, namely a set of $N$ capillary tubes. A fraction $S_{w,i}$ of 
these capillaries has an inner area $a_s$, whereas the rest has an inner area $a_l$.  The capillaries
with the larger area contain bubbles as in section \ref{bubbles}.  The smaller capillaries contain only
wetting liquid as in section \ref{cfb}. The fluid contents in the small capillaries is irreducible.  
The seepage velocity in the smaller capillaries is $v_{sw}$. The
wetting and non-wetting seepage velocities in the larger capillaries are the same,
\begin{equation}
\label{para50}
v_n=v_{lw}=v_l\;.
\end{equation}   
Hence, the seepage velocity is then
\begin{equation}
\label{para51}
v=S_{w,i} v_{sw}+(1-S_{w,i})v_l\;.
\end{equation}
We have that
\begin{equation}
\label{para52}
\frac{dv}{dS_w}=\frac{dv_l}{dS_w}\;,
\end{equation} 
since $v_{sl}$ is independent of $S_w$.
Equation (\ref{eqn10-3-e}) gives
\begin{equation}
\label{para53-1}
\hat{v}_n=S_{w,i} v_{sw}+\left(1-S_{w,i}\right)\left(v_l-S_w\frac{dv_l}{dS_w}\right)\;,
\end{equation}
where we have used (\ref{para51}).
We now use equations (\ref{eqn10-5.6}) and (\ref{para50}) to find
\begin{equation}
\label{para53}
\hat{v}_n-v_n=\hat{v}_n-v_l=-S_w v_m\;.
\end{equation}
Combining equations (\ref{para52}), (\ref{para53-1}) and (\ref{para53}) gives 
\begin{equation}
\label{para54}
v_m=\frac{S_{w,i}}{S_w}\left(v_l-v_{sw}\right)+\frac{dv}{dS_w}\;.
\end{equation}
Hence, the co-moving velocity is a sum of the co-moving velocities of the two previous 
sections, see (\ref{para22}) and (\ref{para32}).   

\begin{figure*}
\includegraphics[width=0.75\textwidth,clip]{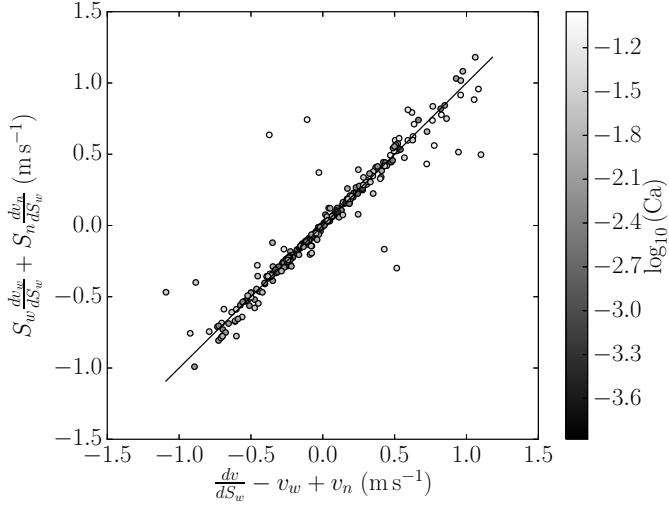}
\caption{The co-moving velocity $v_m$ may be expressed as in equation (\ref{eqn18-1}): 
$v_m=v'-v_w+v_n$, or as in equation (\ref{eqn20-1}): $v_m=S_w v_w'+S_n v_n'$.  
We plot here the two different expressions for $v_m$ against each other. 
The data points are shaded according to the capillary number Ca of the flow. The data  
are based both on the hexagonal and the square lattice.}
\label{fig4}   
\end{figure*}

We check the consistency of this calculation by using the co-moving velocity 
found in (\ref{para54}) together with
equation (\ref{eqn11-5.5}) to calculate $v_w$.  We find as expected
\begin{equation}
\label{para33-1}
v_w=\frac{1}{S_w}\left[S_{w,i} v_{sw}+(S_w-S_{w,i}) v_{l}\right]\;.
\end{equation}
\bigskip

These four analytically tractable examples have allowed us to demonstrate in detail how
the thermodynamic formalism that we have introduced in this paper works.  We have in particular
calculated the co-moving velocity $v_m$ in all four cases.  As is
clear from these examples, it does {\it not\/} depend on the constitutive equation or any other
equation except through $v$ and $dv/dS_w$.   

Hence, the discussion of these four models has not included the constitutive 
equation (\ref{conc-7}).  In order to have the seepage velocities $v_w$ and $v_n$ as
functions of the driving forces, we would need to include the constitutive equation and
the incompressibility condition (\ref{conc-1.1}) in the analysis.  

\section{Network model studies}
\label{numerical}

Our aim in this section is to map the co-moving velocity $v_m$ for a network 
model which we describe in the following.  We compare the measured seepage
velocities with those calculated using the formulas derived earlier.  We use
the theory to predict where the three cross points defined and discussed in section
\ref{miscibility} coincide and verify this prediction through direct numerical calculation. 

\begin{figure*}
\includegraphics[width=1.0\textwidth,clip]{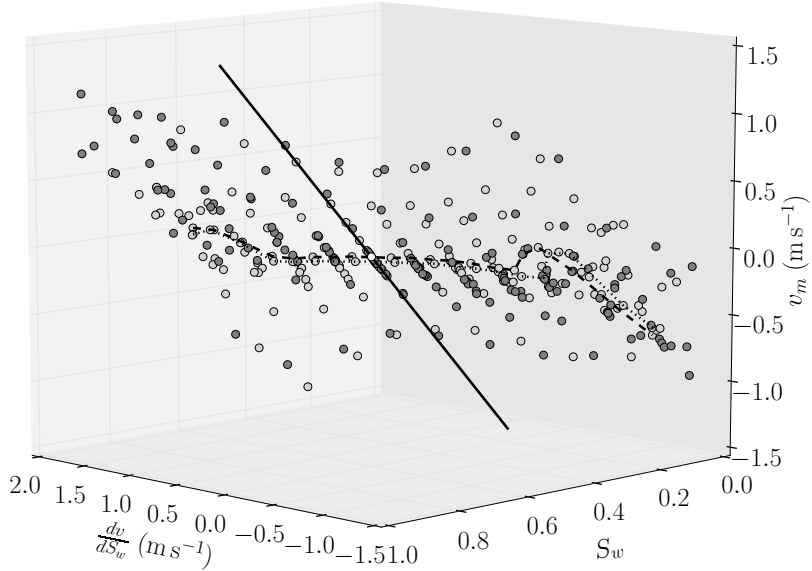}
\caption{Plot of $v_m$ as a function of $S_w$and $dv/dS_w$ for a number 
of different flow parameters for both the hexagonal lattices (light gray data points) and the 
square lattices (dark gray data points).  The point $(s_w,v_m,v')
=(0.6,0,0)$ is highlighted as a white circle. The straight line is $0.79 v'$.
making $b=0.79$ in equation (\ref{eqn25-1}).  The dotted and the dashed curves
correspond to the two data sets shown in figure \ref{fig7}: The curve passing
through the point $(S_w,v_m,v')=(0.6,0,0)$ corresponds to the data set used for 
figure \ref{fig7}b, whereas the other curve is the data set corresponding to that of 
figure \ref{fig7}a.}      
\label{fig5}   
\end{figure*}

The network model first proposed by Aker et al.\ \cite{amhb98} has been 
refined over the years and is today a versatile model for immiscible 
two-phase flow under steady-state conditions due to the implementation
of bi-periodic boundary conditions.  The model 
tracks the interfaces between the immiscible fluids by solving the Kirchhoff 
equations with a capillary pressure in links created by the interfaces they contain
due to a surface tension $\sigma$.  The links are hour glass shaped with average
radii $r$ drawn from a probability distribution.  The minimum size of the bubbles 
of wetting or non-wetting fluid in a given link has a minimum length of $r$, the average 
radius of that link.     

Our network and flow parameters were chosen as follows:  we used hexagonal
(honeycomb) lattices of size $60\times 40$  and square lattices of size $80\times 30$, 
both with all links having the same length $l=1$ mm.  
The radii $r$ of the links were assigned from a flat distribution in the interval
$0.1 l < r < 0.4l$.  The surface tension $\sigma$ was either $0.02$, $0.03$ or 
$0.04$ N/m. The viscosities of the fluids $\eta_w$ and $\eta_n$ were set to either
either $10^{-3}$,  $2\times 10^{-3}$, $5\times 10^{-3}$  and
$10^{-2}$ Pa s.  The pressure drop over the width of the network, $|\Delta P/L|$  
was held constant during each run and set to either 0.1, 0.15, $0.2$ 0.3, 0.5 or 
1 M Pa/m. The capillary number Ca ranged from around $10^{-3}$ to $10^{-1}$, see 
figure \ref{fig4}. 

We measure the seepage velocities $v_w$ as follows. In both
networks, we define layers normal to the flow direction. 
Each layer has an index $j$ and is the set of links intersected by a 
plane normal to the flow direction. In layer $j$, each link has an index $i$. 
For every time step, we calculate
\begin{equation}
\label{eqn25-1}
v_w = \frac{\sum_j \sum_i q_{w,ij}}{\sum_{j'} \sum_{i'} S_{w,i'j'} a_{i'j'}}\;,
\end{equation}
where $S_{w,ij}$ is the saturation of link $i$ in layer $j$ and
$a_{w,ij}$ is the cross-sectional area of the link, projected into the
plane normal to the flow direction. The flow rate $q_{w,ij}$ is
measured as the volume of wetting fluid that has flowed past the
middle of the link during the time step, divided by the length of the
time step. The time-averaged value of $v_w$ is then calculated by
averaging over all time steps, weighted by the time step length. The
non-wetting fluid velocities $v_{n}$ are measured in the same way.

We show in figure \ref{fig4} $v_m=v'-v_w+v_n$ (equation (\ref{eqn18-1})
vs.\ $v_m=S_w v_w'+S_n v_n'$ (equation (\ref{eqn20-1})).  This verifies the
main results of section \ref{seepage}, besides being a measure of the 
quality of the numerical derivatives based on central difference for the 
internal points. Forward/backward difference was used at the end-points 
of the curves where $S_w=0$ or $S_w=1$.  

\begin{figure*}
\includegraphics[width=1.0\textwidth,clip]{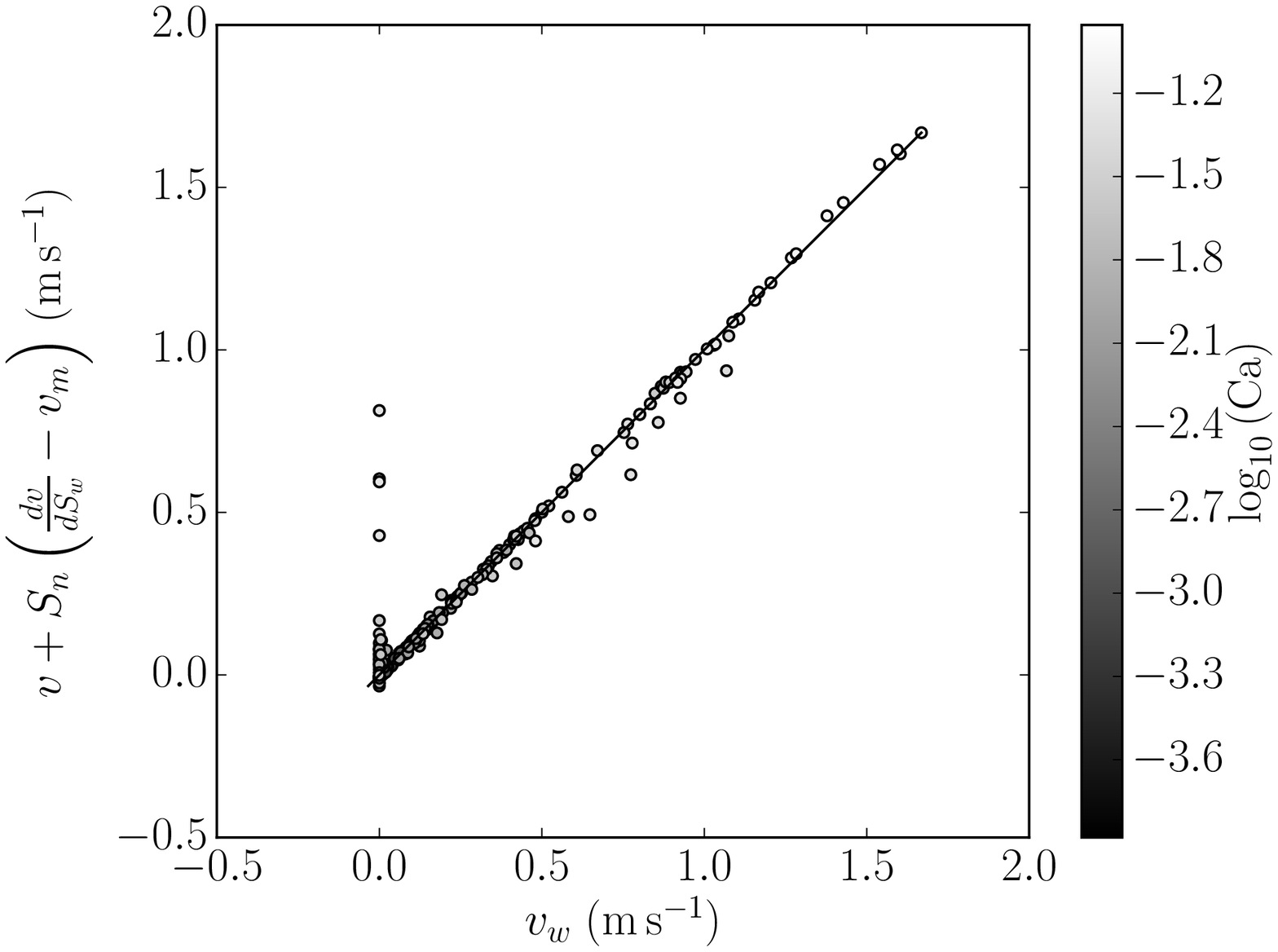}
\includegraphics[width=0.825\textwidth,clip]{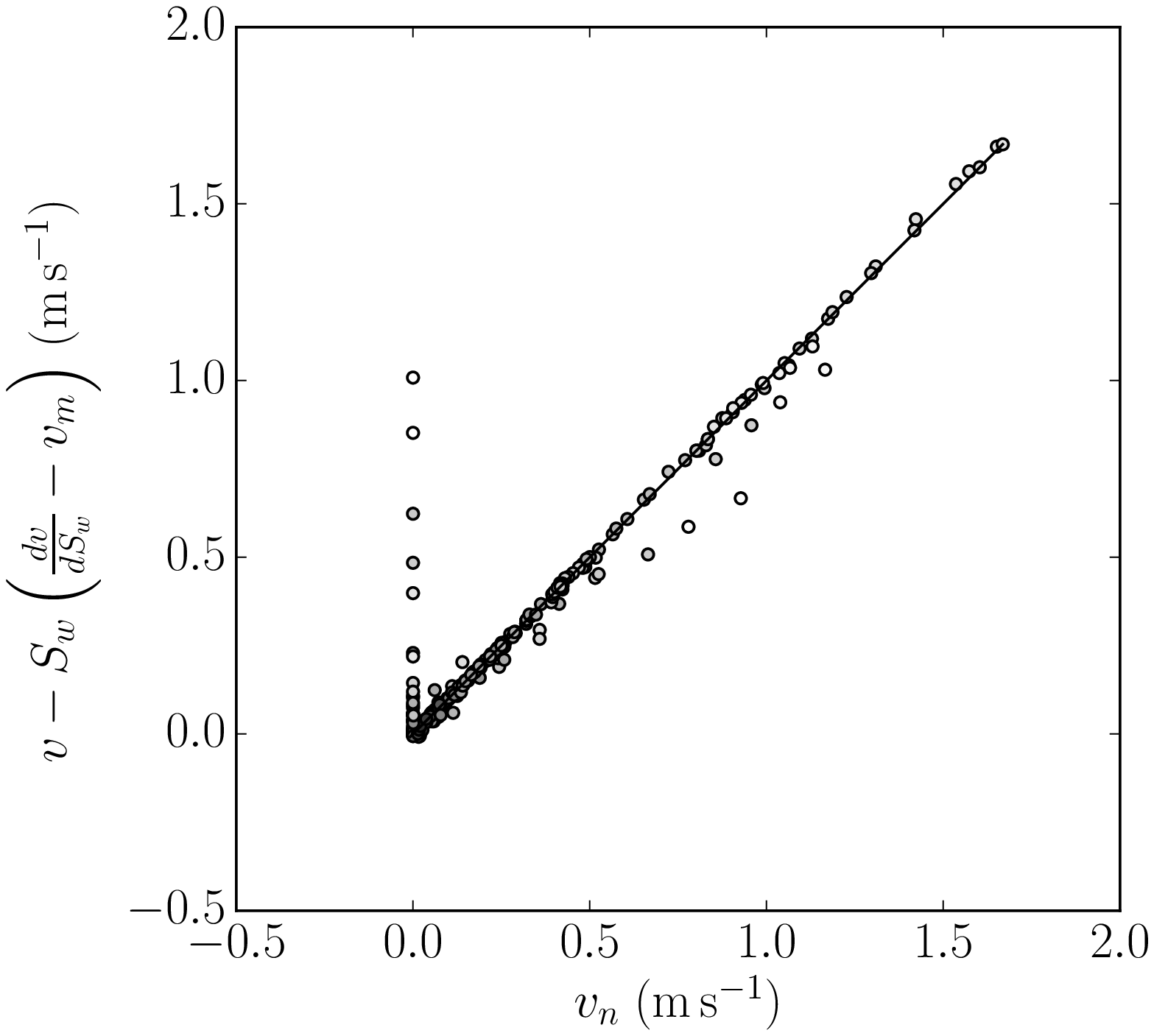}
\caption{We show the seepage velocities $v_w$ (a) and $v_n$ (b)
calculated from equations (\ref{eqn11-5.5}) and (\ref{eqn11-5.6})
vs.\ the measured $v_w$ and $v_n$. The data points are shaded according to 
the capillary number Ca of the flow and are based on both the hexagonal and
the square lattices.}
\label{fig6}   
\end{figure*}

Figure \ref{fig5} shows the co-moving velocity $v_m$ calculated from equation (\ref{eqn20-1})
as a function of $S_w$ and $v'$ for the flow parameters given above.  The data are 
roughly consistent with a linear form 
\begin{equation}
\label{num1}
v_m=aS_w+b\ \frac{dv}{dS_w}-c\;,
\end{equation}
where $a\approx -0.15$, $b\approx 0.79$ and $c\approx -0.095$.  We note that we 
show both the data from the hexagonal and the square lattices in figure \ref{fig5}.  
The co-moving velocity $v_m$ appears not to be sensitive to the lattice topology.      

We compare in figure \ref{fig6} the measured seepage velocities $v_w$ and $v_n$ to the 
seepage velocities calculated from equations (\ref{eqn11-5.5}) and (\ref{eqn11-5.6}) where
we have used the co-moving velocity shown in figure \ref{fig5}.  The spikes at low $v_w$ and $v_n$
are due to the transition from both fluids moving to only one fluid moving creating a jump in
the derivative.     

We note that when the wetting saturation $S_w$ is equal to $S_{w,0}=c/a$ in equation (\ref{num1}), 
the co-moving velocity $v_m$ takes the form $v_m=b v'$.  For the flows we study here, we find $S_{w,0}
\approx 0.6$. With reference to the discussion of the cross points in section \ref{miscibility}, we then 
have $v_m=0$ and $v'=0$ occurring for the same wetting saturation.  Hence, we have $S_w^A=S_w^B=S_w^C$
at this point.

\begin{figure*}
\includegraphics[width=1.0\textwidth,clip]{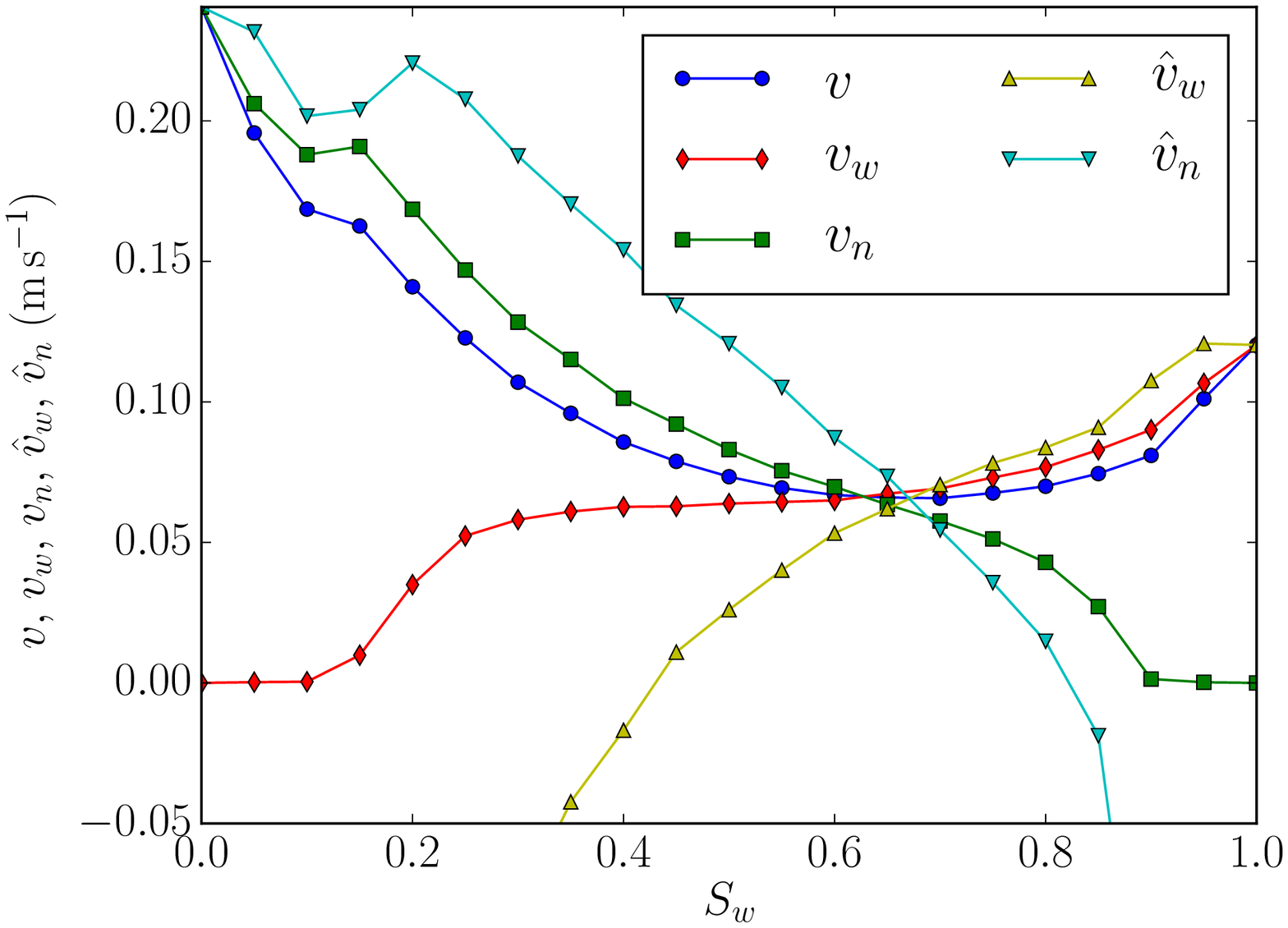}
\includegraphics[width=1.0\textwidth,clip]{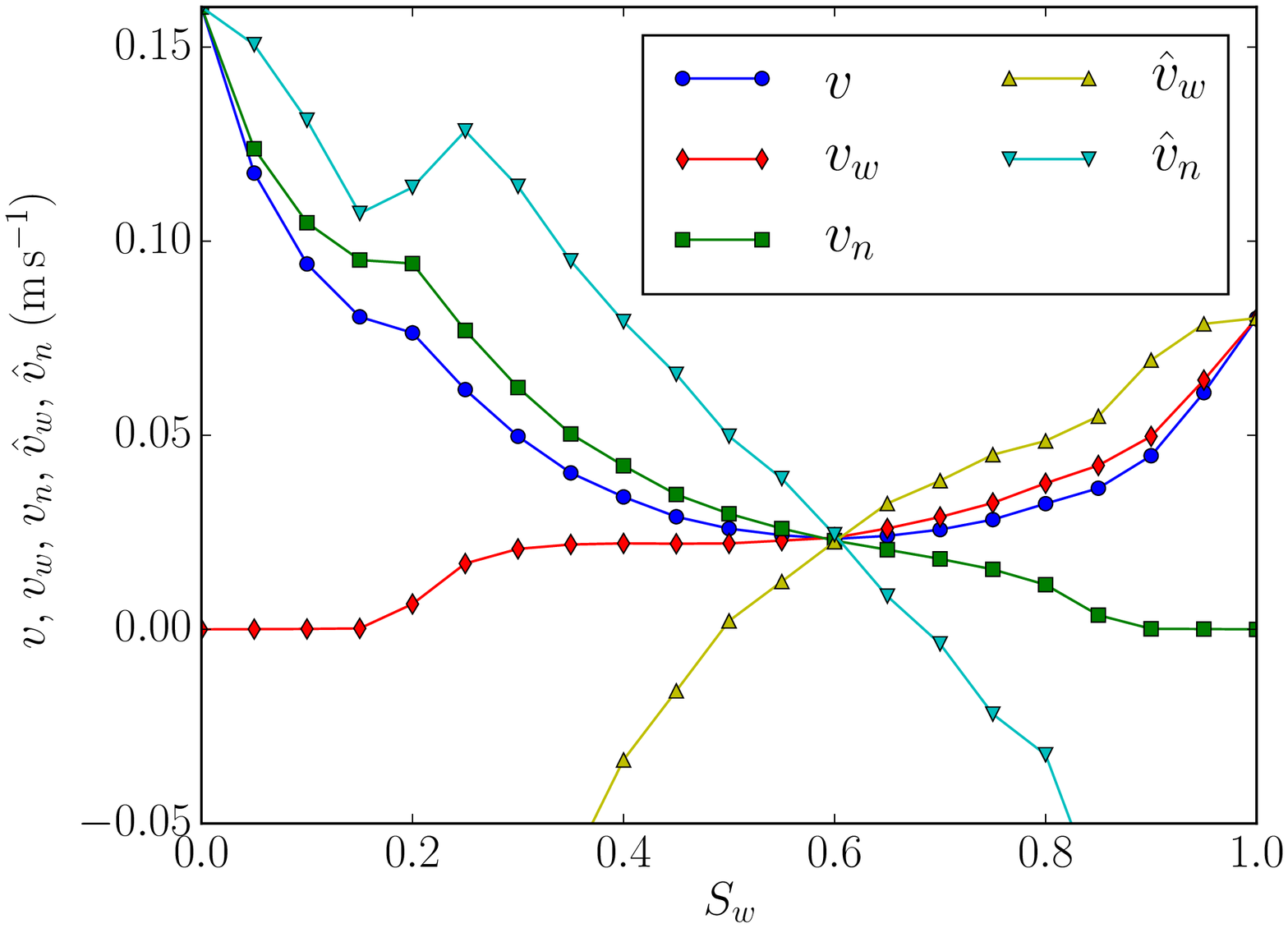}
\caption{Plot of $v$, $v_w$, $v_n$, $\hat{v}_w$ and $\hat{v}_n$ vs.\ $S_w$ for the hexagonal 
lattice.  
In (a), we have set $\Delta P/L$=-0.15 M Pa/m, $\eta_w=0.001$ Pa s, $\eta_n=0.002$ Pa s
and $\sigma=0.03$ N/m. The data set corresponds to the dotted curve in figure \ref{fig5}
for which $v_m=0$ does not coincide with $v'=0$.  We see that the three cross 
points defined in section \ref{miscibility} are ordered according to the inequality 
(\ref{mp2}), $S_w^C\le S_w^B\le S_w^A$.  
In (b) we have set $\Delta P/L$=-0.10 M Pa/m, 
$\eta_w=0.001$ Pa s, $\eta_n=0.002$ Pa s and $\sigma=0.03$ N/m. The data set 
corresponds to the dashed curve in figure \ref{fig5} passing through $(v_m,v')=(0,0)$. 
In this case we find that the cross points all coincide.  Hence, we have $S_w^C= S_w^B= S_w^A$,
as predicted.}
\label{fig7}   
\end{figure*}

Figure \ref{fig5} shows two closely spaced curves, one dotted and one dashed. The dashed
curve passes through the point 
$(v_m,v')=(0,0)$ (marked by a non-filled circle).  We show in figure \ref{fig7} 
$v$, $v_w$, $v_n$, $\hat{v}_w$ and $\hat{v}_n$ corresponding to these two curves. The
thermodynamic velocities have been calculated using equations (\ref{eqn10-2-e}) and
(\ref{eqn10-3-e}). In 
\ref{fig7}a, the flow parameters have been chosen so that $v'=0$ and $v_m=0$ do not coincide as 
we see in figure \ref{fig5}.  The
ordering of the cross points then follow the inequality (\ref{mp2}): $S_w^C\le S_w^B\le S_w^A$.  In figure
\ref{fig7}b, we have chosen the parameters so that $v'=v_m=0$ for a given saturation, $S_w=S_{w,0}
\approx 0.6$, see figure \ref{fig5}.  The theory then predicts $S_w^C= S_w^B= S_w^A$, 
which is exactly what we find numerically. This constitutes a non-trivial test of the theory: from
an experimental point of view, this is the point where $v_w=v_n=v$ {\it and\/} $v$ has its
minimum.  These are all measurable quantities.

\section{Discussion and conclusion}
\label{conclusion}

We have here introduced a new thermodynamic formalism to describe immiscible two-phase flow in 
porous media.  This work is to be seen as a first step towards a complete theory based on
equilibrium and non-equilibrium thermodynamics.  In fact, the only aspect of thermodynamic
analysis that we have utilized here is the recognition that the central variables are Euler
homogeneous functions.  This has allowed us to derive a closed set of equations in section
\ref{non-eq}.  This was achieved by introducing the 
{\it co-moving velocity\/} (\ref{conc-8}).  This is a velocity function that relates
the seepage velocities $v_w$ and $v_n$ to the thermodynamic velocities
$\hat{v}_w$ and $\hat{v}_n$ defined through the Euler theorem in section
\ref{scaling}.  

The theory we have developed here is ``kinetic" in the sense that it only concerns
relations between the velocities of the fluids in the porous medium and not  
relations between the driving forces that generate these velocities and the velocities.
We may use the term ``dynamic" about these latter relations.       

The theory rests on a description on two levels, or scales. The
first one is the pore scale.  Then there is REV scale.  This scale is much
larger than the pore scale so that the porous medium is differentiable. At
the same time, we assume the REV to be so small that we may assume that the
flow through it is homogeneous and at steady state.  At scales much larger than the 
REV scale, we do not assume steady-state flow.  Rather, saturations and
velocity fields may vary both in space and time. It is at this scale that the 
equations in section \ref{non-eq} apply.

The necessity to introduce the co-moving velocity $v_m$ is surprising at first. 
However, it is a result of the homogenization procedure when going from the pore
scale to the REV scale: The pore geometry imposes restrictions on the flow which
must reflect itself at the REV scale.  

One may at this point ask what has been gained in comparison to the relative
permeability formalism?  The relative permeability approach consists in making 
explicit assumptions about the functional form of $v_w$ and $v_n$ through the generalized 
Darcy equations \cite{b72}.
The relative permeability formalism thus reduces the immiscible two-phase
problem to the knowledge of three functions $k_{r,w}=k_{r,w}(S_w)$,
$k_{r,n}=k_{r,w}(S_n)$ and $P_c=P_c(S_w)$.  However, strong assumptions
about the flow have been made in order to reduce the problem to these
three functions; assumptions that are known to be at best approximative.

Our approach leading to the central equations in section \ref{non-eq},
does {\it not\/} make any assumptions about the flow field beyond the 
scaling assumption in equation (\ref{eqn10}).  Hence, the constitutive
equation (\ref{conc-7}) can have any form, for example the non-linear
one presented in \cite{tkrlmtf09,tlkrfm09,rcs11,sh12,sbdkpbtsch16}. 
Hence, our approach is in this sense 
much more general than the relative permeability one.  

\begin{acknowledgements}
The authors thank Carl Fredrik Berg, Eirik Grude Flekk{\o}y, 
Knut J{\o}rgen M{\aa}l{\o}y, Tho\-mas Ramstad, Isha Savani, Per Arne Slotte, 
Marios Valavanides and Mathias Winkler for interesting discussions on this topic.   
AH and SS thank the Beijing Computational Science Research Center CSRC for 
financial support and hospitality. SS was supported by National Natural 
Science Foundation of China under grant number 11750110430. 
This work was partly supported by the Research Council of Norway through 
its Centres of Excellence funding scheme, project number 262644.
\end{acknowledgements}

\end{document}